\documentclass[referee,a4paper,12pt,traditabstract]{swsc} 

\usepackage{graphicx}
\usepackage{txfonts}
\usepackage{subfigure}
\usepackage{epstopdf}
\usepackage[displaymath,mathlines]{lineno}
\usepackage[authoryear,round]{natbib}
\usepackage[backref]{hyperref}
\usepackage{url}

\hypersetup{colorlinks=true,citecolor=cyan,urlcolor=cyan,linkcolor=blue}

\graphicspath{{figs/},{./}}

\newcommand{\dv}[2]{\frac{{\mathrm d}#1}{{\mathrm d}#2}}
\newcommand{\pdv}[2]{\frac{\partial #1}{\partial #2}}
\newcommand{\ppd}[2]{\frac{\partial^2 #1}{\partial {#2}^2}}

\def\degdot{\hbox{$.\!\!^\circ$}}

\newcommand{\revone}[1]{{#1}}

\newcommand{\revtwo}[1]{{#1}}

\begin{document}


\title{Towards an algebraic method of solar cycle prediction} 
\subtitle{I. Calculating the ultimate dipole contributions of individual 
active regions}

\titlerunning{Algebraic method of solar cycle prediction I.}
\authorrunning{Petrovay, Nagy and Yeates}

\author{Krist\'of Petrovay\inst{1} \and Melinda Nagy\inst{1}
\and 
Anthony R. Yeates\inst{2}
}

\institute{E\"otv\"os Lor\'and University, Department of Astronomy, Budapest,
Hungary\\
\email{\href{mailto:K.Petrovay@astro.elte.hu}{K.Petrovay@astro.elte.hu}, 
\href{mailto:M.Nagy@astro.elte.hu}{M.Nagy@astro.elte.hu}
}
\and
Department of Mathematical Sciences, Durham University, Science Laboratories,
Durham, UK\\
\email{\href{mailto:anthony.yeates@durham.ac.uk}{anthony.yeates@durham.ac.uk}}
}



  \abstract
   {         
We discuss the potential use of an algebraic method to compute the
value of the solar axial dipole moment at solar minimum, widely
considered to be the most reliable precursor of the activity level in
the next solar cycle. The method consists of summing up the ultimate
contributions of individual active regions to the solar axial dipole
moment 
\revtwo{at the end of the cycle}. 
A potential limitation of the approach is its dependence on
the underlying surface flux transport (SFT) model details.  We
demonstrate by both analytical and numerical methods that the factor
relating the initial and ultimate dipole moment contributions of an
active region displays a Gaussian dependence on latitude with
parameters that only depend  on details of the SFT model through the
parameter $\eta/\Delta_u$ where $\eta$ is supergranular diffusivity and
$\Delta_u$ is the divergence of the meridional flow on the equator. In a
comparison with cycles simulated in the 2$\times$2D dynamo model we
further demonstrate that the inaccuracies associated with the
algebraic method are minor and the method may be able to reproduce the
dipole moment values in a large majority of cycles.
}

\keywords{solar cycle --
   rogue sunspots --
   surface flux transport modeling
  }

\maketitle


\section{Introduction}

Predicting the amplitude of an upcoming solar cycle is the central
issue of space climate forecasting. It is widely accepted that the
best performing, physically well motivated prediction method is based
on the good linear correlation between the solar axial dipole moment
in solar activity minimum and the amplitude of the next cycle
(\citealt{Schatten_:polar.prec},
\revone{\citealt{Wang_Sheeley:geomg.precursor},
\citealt{Munozjara+:polarfac.precursor}, } 
\citealt{Hathaway_Upton2016}, 
\citealt{Petrovay:LRSP2}). In order to extend the rather short, 3--4
year long time span of these forecasts, in recent years efforts have
been made to ``predict the predictor'', i.e. to model and forecast the
evolution of the magnetic flux distribution over the solar surface 
\revone{(\citealt{Jiang+:1cycle})}.

Starting with the pioneering work of \cite{DeVore1985_firstSFT}, the
standard approach to this problem has been the use of {\it
surface flux transport} (SFT) simulations. This approach assumes that
the line-of-sight magnetic field component shown in synoptic maps
constructed from solar magnetograms corresponds to the projection of
an inherently radial mean photospheric magnetic field with strength
$B_r$, the transport of which is governed by advection due to large
scale flows and diffusion due to supergranular motions:
\begin{eqnarray}
  \pdv{B_r}t &=&\displaystyle  -\Omega(\lambda)\,\pdv{B_r}\phi 
  -\frac{1}{R\cos{\lambda}}\,
  \pdv{}\lambda\left[B_r\,u(\lambda)\cos{\lambda}\right] \nonumber \\
 &&  +\frac{\eta}{R^2}\left[ \frac 1{\cos\lambda}\, 
 \pdv{}\lambda\left(\cos{\lambda}\,\pdv{B_r}\lambda\right) 
 +\frac 1{\cos^2\lambda}\,
 \ppd{B_r}\phi \right] -\frac{B_r}{\tau} + S_r(\lambda,\phi,t) ,
\label{eq:SFT}
\end{eqnarray}
where $t$ is time, $\lambda$ and $\phi$ are heliographic latitude and
longitude, $R$ is the radius of the Sun, $\Omega$ is the angular
velocity of differential rotation, $u(\lambda)=u_0 f(\lambda)$ is the
meridional flow, and $\eta$ is the supergranular diffusivity. The
source term $S_r$ represents the emergence of new flux into the
atmosphere in active regions, while the term $-B_r/\tau$ is the simplest
(though not the most realistic) form of a sink term
representing the effects of radial diffusion that would appear in a
consistent derivation of the transport equation from the radial
component of the induction equation. (For further discussion of the
decay term see e.g. \citealt{Baumann+:decayterm},
\citealt{Whitbread+:SFToptimization} or \citealt{Petrovay+Talafha}.
For general reviews of the SFT modelling approach see
\citealt{Sheeley:LRSP}, 
\revone{\citealt{Mackay+Yeates}, }
\citealt{Jiang+:SFTreview}, \citealt{Wang:SFTreview}).

Numerous studies have been performed with the objective to reproduce
and predict the evolution of the solar dipole moment
(\citealt{Wang_Sheeley_dipmom}, \citealt{Whitbread+:SFToptimization},
\citealt{Virtanen+:SFT}). Despite some spectacular
successes, this approach still suffers from two main limitations:

(1) Ill constrained SFT model ingredients. The model has at least 3
free numerical parameters ($u_0$, $\eta$, $\tau$) as well as a free
function, the observationally not well mapped meridional flow profile
$f(\lambda)$. Attempts to determine the ingredients from direct
observations have dubious relevance and often yield contradictory
results. Internal optimizations of the model, in turn, result in
parameter combinations that may depend on the choice of merit and
still allow a rather wide freedom in the choices (\citealt{Lemerle1},
\citealt{Virtanen+:SFT}, \citealt{Whitbread+:SFToptimization}, 
\citealt{Petrovay+Talafha}). The ill constrained ingredients imply
that in order to obtain a realistic estimate of the uncertainities in
the predictions an ensemble of models with varying parameters should 
be studied; with the need to numerically solve the partial differential
equation (1) for each of these models the process can get rather
lengthy and cumbersome.

(2) A realistic representation of the active region source. These
sources are often represented as simple bipoles instantaneously
introduced into the simulation, or, more realistically, by
assimilating actual magnetograms taken at the time of their central
meridian passage. In reality, however, during the evolution of an
active region its magnetic flux distribution keeps changing as a
consequence of the emergence of new flux and the proper motions of
sunspots and other magnetic flux concentrations driven by subsurface
dynamics and/or by random, localized photospheric flows ---effects
which are not accounted for in the SFT model. The choice of the proper
form of the source and the time of its introduction is therefore a
highly nontrivial problem. And the very large number (thousands) of
active regions arising in a solar cycle makes this task challenging
even in the simplest, idealized case of bipole representation. These
problems are even further aggravated in the case of historical data,
when the objective is to understand and reproduce the course of
evolution of solar activity in centuries past.

The objective of this paper is to consider ways to alleviate these
difficulties by simplifying the prediction method to its bare
essentials. Specifically, in Section~2 we will show that the SFT
modelling approach of solving a partial differential equation can be
reduced to the calculation of an algebraic sum. Sections~3 and 4
demonstrate that the result of this procedure only depends on two
numerical combinations of the model  parameters without the need to
specify the choice of any unknown function. Then, in
Section~\revtwo{5} 
we validate the approach in a comparison with activity cycles
simulated in a dynamo model. These results may pave the way towards a
more robust and effective approach to solar cycle prediction.

\begin{figure}
\begin{center}
\includegraphics[width=\textwidth]{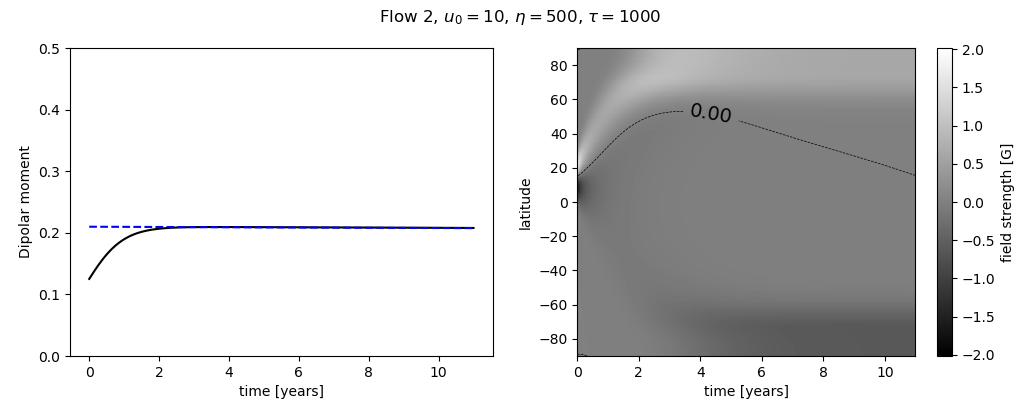}\\
\includegraphics[width=\textwidth]{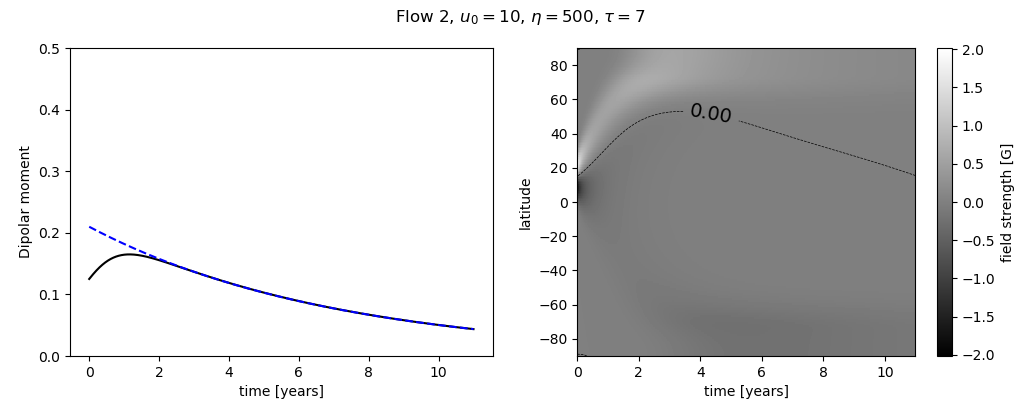}
\end{center}
\caption{Solutions of the 1D SFT equation with a bipolar source placed at
$\lambda_0=+15^\circ$ latitude at time $t=0$. Units of the model parameters
shown on top  of the rows are m$/$s, km$^2/$s and year, respectively. The
meridional flow profile is given by eq.~(\ref{eq:deadzone}). Dashed
blue lines in the left-hand plots mark the asymptotic solutions 
$D=\delta D_\infty$ (top) and $D=\delta D_\infty\exp(-t/\tau)$ (bottom).}
\label{fig:singlesource}
\end{figure}

\section{Mathematical formulation of the problem}

The axial dipolar moment of the Sun is given by
\begin{equation}                       
    D(t) = \frac32 \int_{-\pi/2}^{\pi/2} 
    B(\lambda,t)\sin\lambda\cos\lambda\, \mathrm{d}\lambda ,
 \label{eq:dipmom}
\end{equation}
where $B=\frac 1{2\pi}\int_0^{2\pi} B_r \,d\phi$ is the azimuthally
averaged field strength. 
\revone{Throughout this paper, $D$ will denote the global dipole
moment of the whole Sun as defined in equation (\ref{eq:dipmom}),
while $\delta D$ will denote the contribution to $D$ from an
individual active region.}
The evolution of $B$, and consequently $D$,
is determined by the 1D surface flux transport equation obtained by
azimuthally integrating equation~(\ref{eq:SFT}):
\begin{equation}
  \pdv Bt =\displaystyle -\frac{1}{R\cos{\lambda}}\,
  \pdv{}\lambda\left[B\,u(\lambda)\cos{\lambda}\right] 
  +\frac{\eta}{R^2} \frac 1{\cos\lambda}\, 
 \pdv{}\lambda\left(\cos{\lambda}\,\pdv B\lambda\right) -\frac{B}{\tau} 
  + S(\lambda,t) ,
\label{eq:SFT1D}
\end{equation}
where $S=\frac 1{2\pi}\int_0^{2\pi} S_r \,d\phi$. 

In this azimuthally averaged representation a tilted bipolar region is
then a pair of flux rings of opposite polarity, appearing as a bipole
source with a finite latitudinal separation in equation
(\ref{eq:SFT1D})

Figure \ref{fig:singlesource} presents solutions of equation
(\ref{eq:SFT1D}) for a particular case where the source is a single
bipole placed at $\lambda_0=+15^\circ$ latitude at time $t=0$. Both
polarities were assumed to initially have a Gaussian profile in
$\lambda$ with a half-width $\sigma_0=6^\circ$. The left-hand panels
show the evolution of the axial dipolar moment. The right-hand panels
display the evolution of $B$ in the time-latitude plane. The cases
shown in the top and bottom rows only differ in the value of $\tau$:
in the first case $\tau$ is effectively infinite (no decay), while in
the second case it is shorter than the solar cycle period. As it is
easy to understand from the structure of eq.~(\ref{eq:SFT1D}), the two
solutions only differ by the presence of an exponential factor
$\exp(-t/\tau)$ when $\tau$ is finite.  

The evolution is determined by the competition of two processes. The
diffusive spreading of the two polarity patches of opposite sign leads
to the cancellation of a large part of the flux originally present,
yet a small fraction of the flux still  manages to reach the Southern
hemisphere where it is transported to the 
\revtwo{South}
pole by the meridional flow. The ``leading'' polarity flux patch, 
situated closer to the equator, gives a larger contribution to the
flux ultimately reaching the 
\revtwo{South} 
pole, so in the final state, a leading polarity patch remains at the 
\revtwo{South}
pole, while a corresponding trailing polarity patch remains at the 
\revtwo{North} pole. While the flux in these patches is a fraction of the
original flux in the region, their high latitudinal separation gives
rise to a dipole moment that, in the case plotted in
Fig.~\ref{fig:singlesource}, exceeds the initial value. In the limit
$\tau\rightarrow\infty$ the dipole moment remains very nearly constant
at a fixed value $\delta D_\infty(\lambda_0)$.
For finite $\tau$ the dipole moment asymptotically behaves as 
$\delta D_\infty\exp(-t/\tau)$.

This implies that if, ignoring the complex details of its structure and
evolution, the $i$th active region in cycle $n$ (starting at time
$t_n$) is represented by a simple dipole instantaneously introduced
into the SFT model at time $t_i$ 
\revone{with an initial dipole moment $\delta D_{1,i}$}, 
the ultimate contribution of all active regions at the end of the
cycle will be given by
\begin{equation}
\Delta D_n\equiv D_{n+1} - D_n = 
  \sum_{i=1}^{N_{\mathrm{tot}}} \delta D_{U,i} =
  \sum_{i=1}^{N_{\mathrm{tot}}} \delta D_{\infty,i} \, e^{(t_i-t_{n+1})/\tau} =
  \sum_{i=1}^{N_{\mathrm{tot}}}  
  {f_{\infty,i}}\, \delta D_{1,i} \, e^{(t_i-t_{n+1})/\tau} ,
\label{eq:cycledipmom}
\end{equation}
where $N_\mathrm{tot}$ is
the total number of ARs in the cycle, $\delta D_U$ is the {\it
ultimate} contribution of an active region to the global dipole moment
at the end of a cycle, and the asymptotic dipole contribution factor is
\begin{equation}
f_\infty=\delta D_\infty/\delta D_1 .
\label{eq:finftydef}
\end{equation}
From equation~(\ref{eq:dipmom}) it is straightforward to show that for
a pointlike bipole consisting of a pair of pointlike polarities with a
small latitudinal separation $d_\lambda$ the initial dipole moment
contribution $\delta D_1$ is given by
\begin{equation}
\delta D_1=\frac 3{4\pi R^2}\, \Phi \, d_\lambda\cos\lambda_0 ,
\label{eq:D1}
\end{equation}
where $\Phi$ is the magnetic flux in the northern flux patch.

Equations (\ref{eq:cycledipmom})--(\ref{eq:D1}) offer a
simple algebraic tool to extend the temporal scope of the polar field
precursor method by ``predicting the precursor'', i.e. computing the
dipole moment built up during a solar activity cycle without the need
to solve the partial differential equation~(\ref{eq:SFT1D}) or
(\ref{eq:SFT}). For the case of cycle 23 this approach has been
already exploited by \cite{Jiang+Baranyi} for one particular SFT model
setup.

Generally, however, this approach is subject to a number of limitations. These
limitations were already outlined in the Introduction above. In more
detail, they are:

(1a) The result depends on details of the SFT model used, mainly
through the $f_{\infty,i}$, and for models with a decay term also
through the exponential factor in (\ref{eq:cycledipmom}).

(1b) As illustrated in Fig.~\ref{fig:singlesource}, the asymptotic
solution is approximated after a transitional period of 2--3 years
only. The contribution of active regions emerging in the last years of
the cycle is therefore not accurately represented by this formula. As,
however, activity in this late phase of the cycle is normally rather
low, this is not expected to be a major limitation in most cases.

(2a) The assumption that active regions may be represented by the
instantaneous introduction of simple bipoles into the simulation is
undoubtedly a strong simplification.

(2b) The number $N_{\mathrm{tot}}$ of terms in the sum, i.e. the
number of bipolar regions contributing to the solar axial dipole
moment can be quite high.

In the present work we focus on issues (1a), (1b) and (2a). Issue (2b)
will be dealt with in a companion paper.

\begin{figure}
\begin{center}
\includegraphics[width=0.5\textwidth]{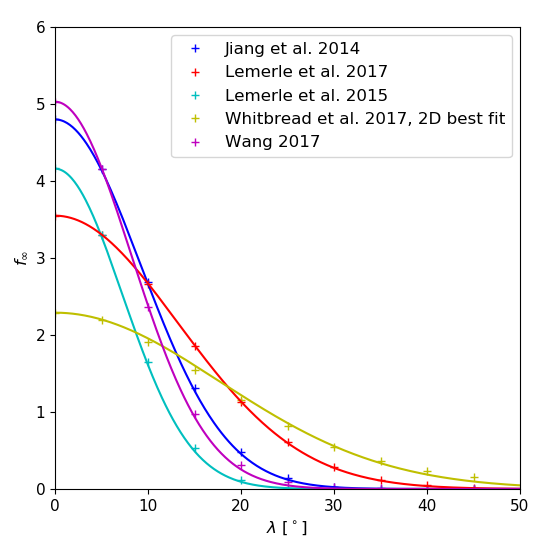}
\end{center}
\caption{Dependence of the \revone{asymptotic} dipole contribution factor
$f_\infty$ of bipolar sources on their latitude in various SFT model
setups. Solid lines are Gaussian fits to the numerical results.}
\label{fig:gaussians}
\end{figure}

\section{Calculating the ultimate dipole contribution of active
regions}

The dependence of the {\it asymptotic dipole contribution factor} $f_\infty$
on latitude was first considered by \cite{Jiang+:scatter}. In a series
of numerical experiments with one particular SFT model they found a
Gaussian dependence on latitude: 
\begin{equation}
f_\infty=A\exp (-\lambda_0^2/2\lambda_R^2 ) .
\end{equation}
In what follows, $\lambda_R$ will be referred to as the {\it dynamo
effectivity range} of active regions.

\begin{figure}
\begin{center}
\includegraphics[width=0.5\textwidth]{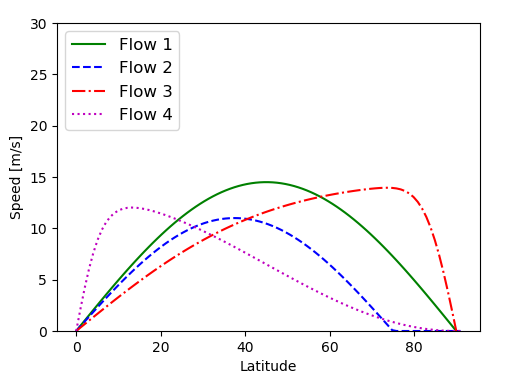}
\end{center}
\caption{The meridional flow profiles used in the paper.}
\label{fig:flows}
\end{figure}

\begin{figure}
\begin{center}
\includegraphics[width=0.48\textwidth]{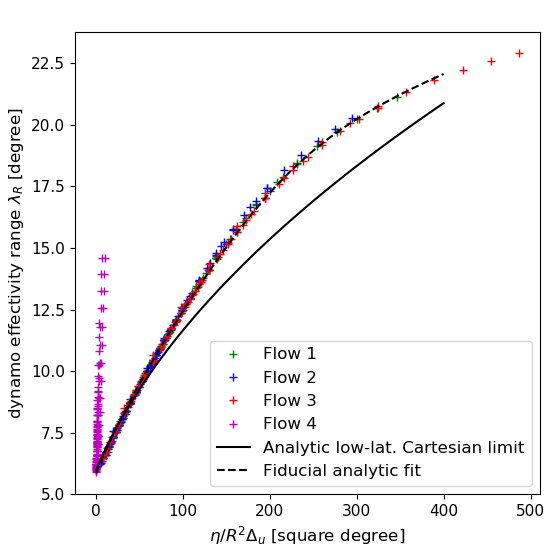}
\includegraphics[width=0.48\textwidth]{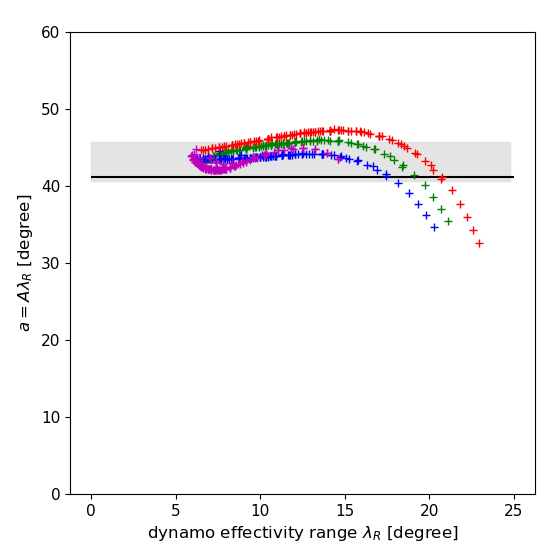}
\end{center}
\caption{Dependence of the dynamo effectivity range $\lambda_R$ (left)
and the amplitude $A$ (right) of the \revone{asymptotic} dipole
contribution factor of bipolar sources on selected parameter
combinations in various SFT model setups for $\sigma_0=6^\circ$. In
the left-hand panel the solid line shows the analytic result
(\ref{eq:lambdaR}) in the low-latitude Cartesian limit, while the
dashed line shows a fiducial analytic fit of the type
(\ref{eq:analfit}). In the right-hand panel the shaded region is the
range expected for a $\sin^n\lambda$ equilibrium polar field profile
with $n>7$; the horizontal line marks the value for $n=8$.}
\label{fig:derange}
\end{figure}

Let us consider whether this conclusion holds generally for other SFT
model setups. In Fig.~\ref{fig:gaussians} the results of a similar set
of experiments for different model setups are shown: in addition
to Jiang et al.'s original setup, the other SFT models used were those
of \cite{Lemerle1} , \cite{Lemerle2},
\cite{Whitbread+:SFToptimization} and \cite{Wang:SFTreview}. It is
apparent that the Gaussian dependence identified by
\cite{Jiang+:scatter} holds generally. Indeed, even in an SFT model
where active region shapes are determined by assimilation rather than
fitted with bipoles, \cite{Whitbread+:dipole} still found a Gaussian.
In all these cases, however, the  width and amplitude of the Gaussian
are different. It was indeed already found by \cite{Nagy+:rogue} that
the dynamo effectivity range in the case of the \cite{Lemerle2} setup
was significantly wider than expected from the results of
\cite{Jiang+:scatter}.

In order to understand the dependence of $\lambda_R$ and $A$ on model
parameters we determine these parameters on a large grid of SFT
models. Our model grid is essentially identical to the grid considered
by \cite{Petrovay+Talafha}, except that here we limit ourselves to
models with effectively no decay term ($\tau=1000\,$years) as the effect
of $\tau$ has already been separated in the exponential factor in
eq.~\ref{eq:cycledipmom}. Four types of flow profiles are considered
(cf.~Fig.~\ref{fig:flows}).

{\it Flow 1:} a simple sinusoidal profile
\begin{equation}
    u=u_{0}\sin(2\lambda  )
\end{equation}

{\it Flow 2:} a sinusoidal profile with a dead zone around the poles,
\begin{equation}
    u =
\left\{
 \begin{array}{ll}
 u_{0}\sin(\pi\lambda/\lambda_{0})  
     & \mbox{if } |\lambda| < \lambda_{0} \\
 0 & \mbox{otherwise } 
 \end{array}
\right.  
\label{eq:deadzone}
\end{equation}

{\it Flow 3:} The \cite{Lemerle2} profile peaking at high latitudes, 
\begin{equation}
\label{eq6} u = \frac{u_{0}}{u_0^\ast}
\mbox{erf}\,(V\cos{\lambda})\, \mbox{erf}\,(\sin{\lambda}) \qquad
u_0^\ast=0.82 \qquad V=7 
\end{equation} 
 
{\it Flow 4:} A profile peaking at very low latitudes, considered by 
\cite{Wang:SFTreview}:
\begin{equation}
u = 1.08\, u_0\, \tanh(\lambda/6^\circ)\,\cos^2\lambda   
\end{equation} 

For each of the profiles, $u_0$ was allowed to vary between 5 and 20
m/s in steps of 2.5, while $\eta$ varied from 50 to 750 km$^2/$s in
steps of 50. From numerical runs like the one plotted in
Fig.~\ref{fig:singlesource} the values of $\lambda_R$ and $A$ were
determined for each model.

Experimenting with different simple combinations of the input and
output parameters we find the clearest relationship in the case
plotted in Fig.~\ref{fig:derange}. Here, 
\begin{equation}
\Delta_u=\frac 1R \left.\dv{u}{\lambda}\right|_{\lambda=0}
\end{equation}
is the divergence of the meridional flow at the equator.

The finding that the single parameter combination $\eta/\Delta_u$
determines $f_\infty$ for all but one of our 2-parameter model grids, 
irrespective of the choice of the meridional flow profile is an
impressive and 
\revone{somewhat} 
unexpected result which calls for a theoretical interpretation. 

\section{Analytical derivation of the \revone{asymptotic} 
dipole contribution factor}

\subsection{Low-latitude Cartesian limit}

To make the problem analytically tractable, we limit ourselves to the
neighbourhood of the equator ($\lambda\ll 1$ radian)  where the
spherical coordinate grid may be locally approximated by a Cartesian
setup and meridional flow  is expressed by the leading term of its
Taylor expansion as $u=R\Delta_u\lambda$ with $\Delta_u=$const. This  is formally
identical to the cosmological case of a one-dimensional Hubble flow in
a vacuum dominated universe and the advection of a frozen-in magnetic
field configuration by the flow is an exponential expansion. This
cosmological analogy suggests to consider the evolution in the
Lagrangian comoving (co-expanding) frame, where fluid elements are
labelled by their latitude $\lambda_L$ at time zero, rather than their
current latitude $\lambda=\mathrm{e}^{\Delta_u t}\lambda_L$.

We recall the initial condition of the evolution, as illustrated in
Fig.~\ref{fig:singlesource}: a pair of opposite polarity flux rings
with Gaussian profile of half width $\sigma_0\ll 1$ radian, situated
at latitude $\lambda_0$, with a separation $d_\lambda\ll 1$ radian
between the rings. (In the diffusive case, other assumed initial
profiles will also soon approach Gaussian by virtue of the central limit
theorem.) What we are looking for is the amount of net transequatorial
flux (flux in the other hemisphere) in the limit $t\rightarrow\infty$.

\subsubsection{Asymptotic magnetic field profile}

In the Lagrangian frame flow advection is absent by definition, so the
flux transport equation simplifies to a diffusion equation 
\begin{equation}
\frac{dB_L}{dt}=\frac{\eta_L}{R^2}\ppd {B_L}{\lambda_L} ,
\label{eq:timedepdiffeq}
\end{equation}
where, however, the diffusivity $\eta_L(t)$ is now time dependent.
Indeed, in the comoving frame the unit of length expands exponentially
as $\exp(\Delta_u t)$, so the diffusivity, of dimension length$^2/$time,
expressed in these units, will scale as $\eta_L\propto e^{-2\Delta_u t}$.
For the same reason, the Lagrangian flux density $B_L$ is related to the
Eulerian by $B=B_L\mathrm{e}^{-\Delta_u t}$.

Consider the evolution of one of the two flux patches comprising the
bipole. Our initial condition is 
\begin{equation}
B_L(\lambda_L,0) = 
  \frac{a}{\sigma_0}\exp -\frac{(\lambda_L-\lambda_0)^2}{2\sigma_0^2}.
\end{equation}

This problem may be solved exactly using Fourier transforms
\citep[cf.][]{Mackay2016}. 
First, we change the time variable from $t$ to
$t'={\textstyle \frac 1{R^2}} \int_0^t\eta_L(\tilde{t})\,\mathrm{d}\tilde{t}$ 
to obtain the standard diffusion equation
\begin{equation}
\frac{\partial B_L}{\partial t'} = \frac{\partial^2 B_L}{\partial\lambda_L^2},
\label{eq:sdiff}
\end{equation}
which may be solved using standard techniques. In particular, if we
define the Fourier transform
\begin{equation}
\hat{B}_L(k,t') = \frac{1}{2\pi}\int_{-\infty}^\infty B_L(\lambda_L,t')
  \mathrm{e}^{-ik\lambda_L}\,\mathrm{d}\lambda_L
\end{equation}
then equation (\ref{eq:sdiff}) implies that $\hat{B}_L(k,t') =
\hat{B}_L(k,0)\mathrm{e}^{-k^2t'}$. The Fourier transform of our initial
condition is
\begin{equation}
\hat{B}_L(k,0) = \frac{a}{\sqrt{2\pi}}\exp-\frac{\sigma_0^2k^2 + 2\lambda_0ki}{2}.
\end{equation}
Inverting the transform finally gives
\begin{equation}
B_L(\lambda_L,t') = \int_{-\infty}^\infty\hat{B}_L(k,t')\mathrm{e}^{ik\lambda_L}\,
  \mathrm{d}k =  \frac{a}{\sqrt{2t' + \sigma_0^2}}
  \exp-\frac{(\lambda_L-\lambda_0)^2}{2(2t' + \sigma_0^2)}.
\end{equation}
Since $t' = -\frac{\eta}{2R^2 \Delta_u}(\mathrm{e}^{-2\Delta_ut}-1)$,
we arrive at
\begin{equation}
B_L = \frac{a}{\sigma(t)}\exp - 
    \frac{(\lambda_L-\lambda_0)^2}{2\sigma(t)^2},
\label{eq:BLsolt}
\end{equation}
with 
\begin{equation}
\sigma(t)=\left[\sigma_0^2+\frac\eta{R^2 \Delta_u}(1-e^{-2\Delta_u t})\right]^{1/2} .
\label{eq:sigma}
\end{equation}

Note that this self-similar solution might have been anticipated,
given that the Gaussian is known to be a self-similar solution of the
diffusion equation and the meaning of $B_L$ is [one-dimensional] flux
density, so magnetic flux conservation requires the amplitude of the
Gaussian to scale with the inverse of $\sigma$. Plugging
eq.~(\ref{eq:BLsolt}) back into (\ref{eq:timedepdiffeq}) then returns
(\ref{eq:sigma}).

\subsubsection{Transequatorial flux}

In the $t\rightarrow\infty$ limit equation (\ref{eq:BLsolt}) gives
\begin{equation}
B_{L,\infty} = \frac{a}{\lambda_R(t)}\exp - 
    \frac{(\lambda_L-\lambda_0)^2}{2\lambda_R(t)^2},
\label{eq:BLsol}
\end{equation}
with
\begin{equation} 
\lambda_R=\lambda_{R,\mathrm{limit}}\equiv 
\left(\sigma_0^2+\frac\eta{R^2 \Delta_u}\right)^{1/2} .
\label{eq:lambdaR}
\end{equation} 

\revone{Using equation (\ref{eq:BLsol}), and taking $\lambda_0>0$
for concreteness, the}
fraction of flux of this single polarity in the opposite 
\revone{(Southern)}
hemisphere is given by
\begin{equation} 
f_\Phi(\lambda_0)=
\int_{-\infty}^0 B_{L,\infty}\,d\lambda_L\left/ \int_{-\infty}^\infty
B_{L,\infty}\,d\lambda_L\right. =
\frac 12\left[1-{\rm erf}\,(\lambda_0/\sqrt 2\lambda_R) \right] .
\label{eq:fluxsingle}
\end{equation} 
For our pair of flux rings separated by $d_\lambda$, 
the net transequatorial flux fraction is then 
\begin{equation} 
\Phi_\infty/\Phi_1 = 
f_\Phi(\lambda_0-d_\lambda/2)-f_\Phi(\lambda_0+d_\lambda/2)\simeq
\frac{d_\lambda}{2^{1/2}\pi^{1/2}\lambda_R}
\exp{\frac{-\lambda_0^2}{2\lambda_R^2}} ,
\label{eq:fluxfraction}
\end{equation} 
where the last form is the leading term in a Taylor expansion for
small $d_\lambda$.

\subsection{Sphericity effects}

To compute the 
\revone{asymptotic}
dipole contribution 
\revone{$\delta D_\infty$ from $\Phi_\infty$}
by equation (\ref{eq:dipmom}) we need to return to spherical geometry.
The poleward meridional flow results in a ``topknot'' equilibrium
field distribution strongly peaked near the poles (\citealt{Sheeley+:topknot}).
Indeed, approximating the field profile as
$B(\lambda)\propto\sin^n\lambda$, even flows mildly concentrated on
the poles will result in $n\simeq 7$ (cf. Fig.~4 in
\citealt{Wang:SFTreview}).
Observational constraints indicate $n\ga 8$--$9$
(\citealt{Petrie:LRSP}). With this approximation, using
eqs.~(\ref{eq:dipmom}), (\ref{eq:finftydef}), (\ref{eq:D1}) and
(\ref{eq:fluxfraction}), we have the following expression for the
\revone{asymptotic} 
dipole contribution factor in the low-latitude limit:
\begin{equation} 
f_\infty=
\frac a{\lambda_R} \exp{-\frac{\lambda_0^2}{2\lambda_R^2}}
\qquad  
\mathrm{with} \qquad a=\left(\frac 2\pi\right)^{1/2} \frac{n+1}{n+2} .
\label{eq:finfty}
\end{equation} 
(A factor $1/\cos\lambda_0$ originating from (\ref{eq:D1}) has been
omitted as in the low-latitude limit it becomes unity.) For $n=8$ this
yields $a=41{\degdot}16$; values for $n>7$ are in the range between
$40{\degdot}66$ and $45{\degdot}7$. The curves in the right-hand
panel of Figure~\ref{fig:derange} are in agreement with this result.

The curves representing flow types 1 to 3 in the left-hand panel of
Fig.~\ref{fig:derange} are well fitted by the solution
(\ref{eq:lambdaR}) at low values of $\lambda_R$. This is to be
expected as these flows peak at latitudes above $40^\circ$ so for low
latitudes the profiles are well approximated as linear. For the same
reason, while for values $\lambda_R\sim 10$--$20^\circ$ curvature
effects do come into play, the curves still do not diverge as 
\revone{the nonlinearity of the flow profile} 
remains low, hence, with an appropriate planar projection of the
spherical surface, the flow may still be transformed out switching to
a homologously expanding Lagrangian frame. In this case equation
(\ref{eq:timedepdiffeq}) generalizes to the diffusion equation on a
spherical surface which has no flux-conserving solution with an
exactly Gaussian profile, though Figure \ref{fig:gaussians} indicates
that at moderate latitudes the solutions may still be well
approximated by a Gaussian. The dynamo effectivity range, however,
changes. A good empirical fit to the curves is found to be
\begin{equation} 
\lambda_{R,\mathrm{fit}}=g^{1/2}(x=\eta/R^2 \Delta_u) \lambda_{R,\mathrm{limit}}
\label{eq:analfit}
\end{equation} 
with
\begin{equation} 
g(x)=(m_1 x+ c_1) \{1-\tanh[(x-c_0)/w]\} +(m_2 x+ c_2)
\tanh[(x-c_0)/w] .
\end{equation} 

The points representing flow type 4 in the left-hand panel of
Fig.~\ref{fig:derange} strongly
diverge from both equation (\ref{eq:lambdaR}) and (\ref{eq:analfit}). The
reason is that for this profile peaking at very low latitude, the
nonlinearity of the profile becomes important already at
$\lambda\sim\sigma_0$. In effect, in most of the area covered by
the AR flux during its evolution the expansion rate will be far below
the nominal equatorial value ---at $\lambda>13^\circ$ the surface will
actually already contract, strengthening the field. Hence, nominal
values of the parameter $\eta/\Delta_u$ are not really relevant for the
determination of $\lambda_R$ in this case.

To close off our analytic discussion we note that the expression
(\ref{eq:lambdaR}) for the dynamo effectivity range can also be
understood in simple physical terms. The time scale associated with
diffusion to the other hemisphere from latitude $\lambda$ is
$(R\lambda)^2/\eta$. The latitude where this equals the advective
time scale $1/\Delta_u$ is just $(\eta/{R^2 \Delta_u})^{1/2}$: for higher
latitudes diffusion cannot compete with the poleward advection and
little flux from here can reach the other hemisphere.

\begin{figure}
\begin{center}
\includegraphics[width=\textwidth]{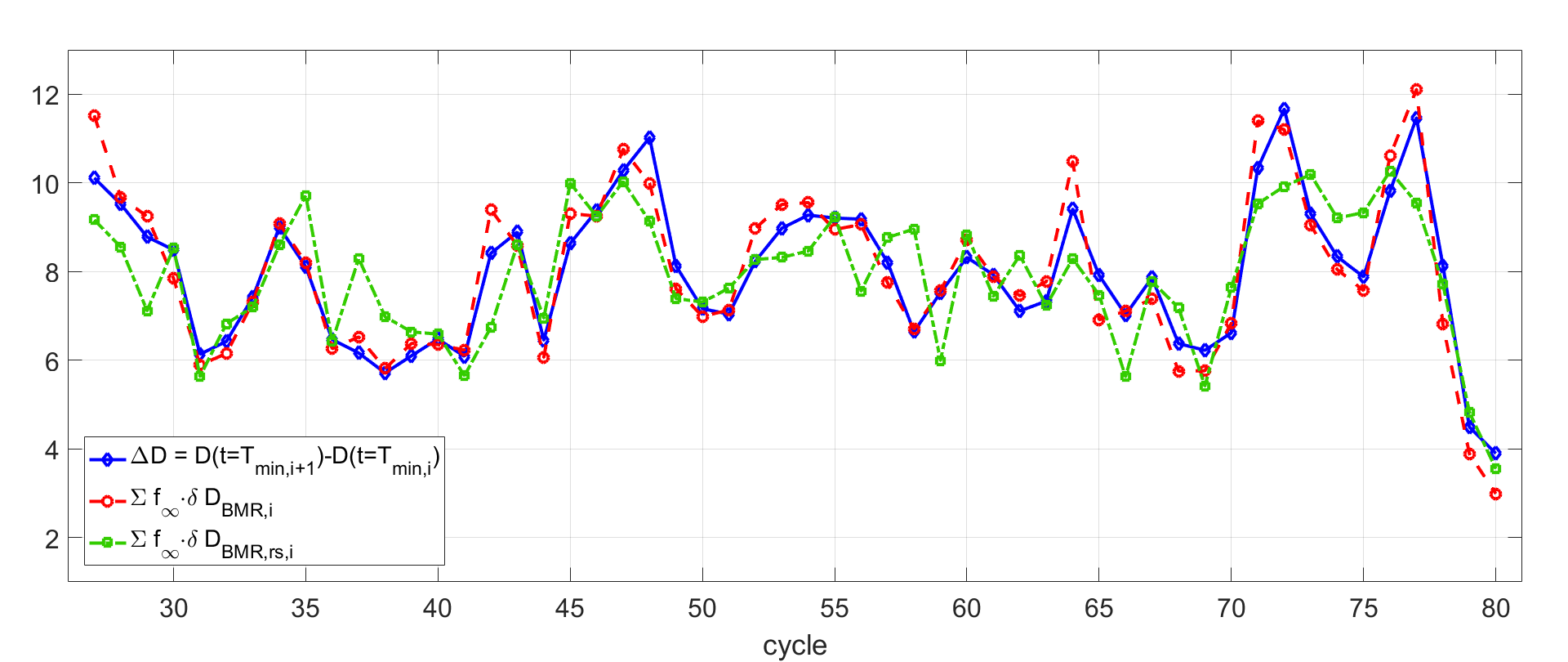}\\
\end{center}
\caption{Comparison of total net change in the solar axial dipole
moment 
\revone{$|D_{n+1}-D_n|$} 
in the 2$\times$2D dynamo simulation and its approximations with the
algebraic method in a segment of 80 cycles. Plotted are absolute
values of the quantities shown in the legend.
See text for further explanations.}
\label{fig:2x2D}
\end{figure}

\section{Comparison to a dynamo model solution}

Our suggested algebraic approach to solar cycle prediction consists in
using equation (\ref{eq:cycledipmom}) to calculate the net dipole
moment at the end of a solar cycle, where the dynamo effectivity
$f_\infty$ is given by eq.~(\ref{eq:finfty}), with $\lambda_R$ and $a$
taken from either a direct interpolation of the numerical results
plotted in Fig.~\ref{fig:derange}, from their analytical approximation
(dashed curve) or even from the low-latitude limit (\ref{eq:lambdaR}).

In order to test the validity and accuracy of the suggested algebraic
approach, in Fig.~\ref{fig:2x2D} we compare the results of a run of
the 2$\times$2D dynamo model, as described in \cite{Lemerle2}, with
the results of our algebraic approach. For the SFT parameters used in
the \cite{Lemerle2} model ($u_0^{\ast}=0.82$,  $u_0=17\,$m$/$s,
$\eta=600\,$km$^2/$s, $\tau=10\,$yr) the numerical results plotted in
Fig.~\ref{fig:derange} give $\lambda_R = 13{\degdot}6$ and $A=3.75$.
The advantage of using the 2$\times$2D model is that it explicitly
includes a 2D SFT model as one of its components, with the source term
represented as idealized bipoles with parameters that can be directly
extracted from the models. In this way, factors like an arbitrary
choice of model parameters or ill-specified sources will not influence
the comparison.

Plotted in blue is the net change in the axial dipole moment 
\revone{$|D_{n+1}-D_n|$}
between subsequent minima of the cycle in the model. The red dashed
curve shows the values computed by adding to the actual dipole moment
at the start of a cycle the expected total dipole moment contribution
calculated by the algebraic method, eqs. (\ref{eq:cycledipmom})  and
(\ref{eq:finfty}), with the parameter values quoted above;
\revone{$\delta D_{1,i}$ values are computed using the BMR properties
extracted from the dynamo code}. 
While the overall agreement is quite good, some smaller discrepancies
remain. The standard deviation of the relative error is 10.1\,\%.
Sources contributing to this unexplained variance include the
invalidity of formula (\ref{eq:cycledipmom}) for active regions
emerging in the last three years of each cycle: as illustrated in
Fig.~\ref{fig:singlesource}, in these cases the time elapsed from
emergence to solar minimum is too short for the asymptotic solution to
set in. Other contributing factors are smaller differences in model
details between the 2$\times$2D model and the 1D SFT model forming the
basis of our algebraic approach. 

The green dashed curve, in turn, diplays the result of the algebraic
method 
\revone{for the ``reduced stochasticity'' case. In this case,  
bipolar magnetic regions (as the active regions are called in the
2$\times$2D model) are substituted with regions of the same flux and
latitude, but with tilt and polarity separation values corresponding
to the expected values for the given flux and latitude, as given by
eqs.~(15) and (16a) in \cite{Lemerle1}.}
While the agreement is noticeably poorer (standard deviation of the
relative error now reaches 21.2\,\%), the net dipole moment change is
still reproduced with less than 25\% error in about 90\,\% of the
cases. This indicates that detailed knowledge of the structure and
evolution of each individual active region may not be indispensable
for a tolerably good predictive skill in the algebraic method in
most (though not all) cycles. This issue will be discussed further in
the second paper of this series (\citealt{Nagy+:algebraic2}).

\section{Conclusions}

We have discussed the potential use of an algebraic method to predict
the value of the solar axial dipole moment at solar minimum. The
method, already applied in the case of one particular SFT model by
\cite{Jiang+Baranyi}, consists in summing up the ultimate
contributions of individual active regions the solar axial dipole
moment as given by equations (\ref{eq:cycledipmom}) and
(\ref{eq:finfty}). 

In Section~2 we listed four potential limitations of the approach.
The first of these, (1a) was its dependence on SFT model details.
\revone{Indeed, the meridional flow profile is still rather uncertain and 
potentially variable; in addition, systematic inflows towards the
activity belts are also expected to impact on flux transport, see
Nagy, Lemerle and Charbonneau (2020, in this issue). Disregarding time 
dependent effects here}
we demonstrated by both analytical and numerical
methods that the dynamo effectivity range $\lambda_R$ and the
equatorial 
\revone{value of the asymptotic} 
dipole contribution factor $A$ only depend on details of the SFT model
through the parameter $\eta/\Delta_u$. This significantly
reduces the uncertainty introduced by the choice of model details and
makes the algebraic method preferable to the more
computation-intensive traditional method of numerically solving the
SFT equation. 

While numerical costs and a more limited freedom in the choice of
parameters advocate the use of this algebraic approach, this clearly
goes at the cost of accuracy. One source of the inaccuracy, (1b), is
inapplicability of the ultimate dipole contribution to active regions
appearing in the last years of a solar cycle, as the contributions of
such ARs have not yet reached equilibrium at the time of minimum. In a
comparison with cycles simulated in the 2$\times$D dynamo model we
demonstrated that the inaccuracy introduced by this effect and by
differences in the underlying numerical models is small and
does not constitute a great obstacle in the way of correctly
reproducing the dipole moment.

Another source of inaccuracy of this approach, (2a), is that the
representation of active region sources by idealized bipoles is likely
to be far from perfect. Applying our method to the same dynamo-simulated
cycles but substituting the assumed initial AR dipole moments with
their expected values for an AR of the given size and latitude showed
that in this case the dipole moment could still be reasonably well
reproduced in the large majority of cycles, lending further support to
the algebraic approach. Nevertheless, in a small fraction of cycles
inaccurate representation of the sources does lead to significant
inaccuracies in the resulting dipole moment values.
Note that while in the dynamo model used here for comparison ARs
were assumed to be be bipolar, in applications to solar data a further
source of uncertainty concerns to what extent a bipolar representation
reflects the structure of ARs (cf.\ \citealt{Iijima+:asym}, 
\revone{\citealt{Jiang+Baranyi},}
\citealt{yeates2020}).

The fourth limitation of the method, (2b) is related to the very high
number of terms in the summation that are theoretically needed for the
correct representation of the dipole contributions, which exacerbates
the issue with the correct representation of the initial AR
contributions. On the other hand, the fact that there are many regions
might help fluctuations from the Gaussian trend to average out in an
overall prediction. This issue will be discussed further in the second
part of this series.

\begin{acknowledgements}

This research was supported by the Hungarian National Research,
Development and Innovation Fund (grant no. NKFI K-128384), by the UK
STFC (grant no. ST/S000321/1) and by the European Union's Horizon 2020
research and innovation programme under grant agreement No. 739500. 
The collaboration of the authors was facilitated by support from the
International Space Science Institute in ISSI Team 474. 

\end{acknowledgements}




\end{document}